\documentstyle[aps,prl,floats,epsf]{revtex}

\newcommand{\bastar}{\begin{eqnarray*}}
\newcommand{\eastar}{\end{eqnarray*}}
\newskip\humongous \humongous=0pt plus 1000pt minus 1000pt

\newif\ifdtup

\relax

\begin{document}
\twocolumn[\hsize\textwidth\columnwidth\hsize\csname@twocolumnfalse%
\endcsname
\title{Model for self-tuning the cosmological constant} 
\bigskip

\author{Jihn E. Kim, Bumseok Kyae and Hyun Min Lee } 
\address{ Department of Physics and Center for Theoretical
Physics, Seoul National University,
Seoul 151-742, Korea \\ \vskip 0.3cm
}
\maketitle

\begin{abstract}
The vanishing cosmological constant in the four dimensional
space-time is obtained in a 5D Randall-Sundrum model with
a brane (B1) located at $y=0$. The matter fields can be located at the
brane. For settling any vacuum energy generated at the
brane to zero, we need a three index antisymmetric tensor 
field $A_{MNP}$ with a specific form for the Lagrangian. 
For the self-tuning mechanism, the bulk cosmological constant 
should be negative.

\vspace{0.3cm}
\end{abstract}

\narrowtext
%\widetext
\bigskip
			]
\def\L{$\Lambda$}
\def\F{$\{F\tilde F\}$}
\def\p{\partial}

The cosmological constant problem is the most serious hierarchy
problem known for more than two decades\cite{veltman}. So far,
there have been several attempts~\cite{attempts} toward understanding this
hierarchy problem, but there has not appeared a fully accepted 
solution yet. The hierarchy in this problem is, $\lq\lq$Why is the scale 
for the cosmological constant \L\ so small compared to the
Planck mass scale $M_P\equiv 2.44\times 10^{18}$~GeV?"
This hierarchy problem has become even more difficult 
with the recent observation of the small but nonvanishing 
vacuum energy of order $(0.003\ {\rm eV})^4$ \cite{perl}. 
The quintessences have been considered
to explain the smallness of this tiny vacuum energy~\cite{quint},
but the bottom line of these ideas is that there exists a solution
of the cosmological constant problem.

The cosmological constant was introduced by Einstein in 
1917 to make the universe static, since it
appeared at that time that the universe seemed to be not evolving.
But the discovery of the expanding universe in 1929 did not
need a static universe, and the cosmological constant has become
another parameter in general relativity. We expect that if a theory
describes physics at the mass scale of order $m$, then
parameters in the theory are expected to be of order $m$.
Gravitation is described at the Planck scale(or inverse Newton's
constant) of order $10^{19}$ GeV. However, the cosmological constant
appearing in the gravity equation is phenomenologically very strongly
bounded $<(0.01\ {\rm eV})^4$, which implies that there is a hierarchy of
order $10^{-120}$ between parameters in the gravity theory.
This hierarchy problem could have been questioned even at the
time of Einstein, not from the static universe condition but
as an hierarchy problem. 

In theoretical physics, the cosmological constant problem has
become a serious one in view of the need for spontaneous symmetry
breaking(SSB) in particle physics~\cite{veltman}, since the vacuum
energy (which is another name for cosmological constant) in SSB is not
fixed by any symmetry principle. The electroweak symmetry breaking
and QCD chiral symmetry breaking can introduce vacuum energies. Thus, the
difficulty of solving the cosmological constant problem in the
four dimensional(4D) space time lies in that the limit \L$\rightarrow 0$
does not introduce any new symmetry. Thus, it may be necessary to go
beyond the 4D space-time or introduce a more general form of the Lagrangian.

In this Letter, we consider a solution of the cosmological constant
problem with one extra dimension. In particular, we work with
one brane located at $y=0$ (B1 brane) where $y$ is the 5th dimension(5D), 
which is the so-called Randall-Sundrum II model (RSII)~\cite{rs2}.
This RSII model is an alternative to compactification of the
extra dimension $y$, but we can obtain an effective 4D flat theory if
gravity is localized at the B1 brane. If we live at B1 and the
graviton wave function is sufficiently damped at large $y$, the
bulk of this extra $y$ space is not of much relevance to us even though
the 5D space is not compactified. Within this kind of setup
there may exist a possibility to attack the cosmological constant
problem again, since the 4D flat space solutions are obtained
with fine tuning(s) even though the 5D cosmological constant is 
nonvanishing and negative. [Note that in 4D there is no possibility
of a flat space solution if the cosmological constant is nonzero.] 

In the RS type models, the idea for self-tuning of the cosmological
constant has been suggested already~\cite{kachru}. We define {\it
the self-tuning model as a model allowing a flat space solution
without fine-tuning between parameters in the Lagrangian.} This
definition is consistent with the one adopted by Hawking and Witten in the
early eighties~\cite{hawking,witten}. They did not care about
the existence of de Sitter or anti de Sitter space solutions,
but needed the existence of a flat space solution with
one undetermined integration constant which can be used to
adjust, given the parameters in the Lagrangian. But in the 4D
space-time theirs was just an idea but not a working model, since 
their 3-form field is not a dynamical field. In the recent
works~\cite{kachru}, the self-tuning idea has been revived in
5 dimensional space-time. Arkani-Hamed et al and Kachru et
al~\cite{kachru} used a specific potential for a bulk scalar
field in the RSII model~\cite{rs2} and obtained a flat space 
solution for a finite range
of input parameters in the Lagrangian. Thus, it seemed that
they realized the self-tuning idea. However, their solution
contained an essential singularity, whence they just
cut off the bulk for a finite 4D Planck mass.
But one cannot ignore this
singularity. If we ignore it, we would not obtain a vanishing
effective cosmological constant in an effective 4D theory,
which contradicts with the flat space solution. 
If the cure for this singularity is correctly performed,
indeed the effective 4D cosmological constant is zero
but one needs one fine-tuning~\cite{nilles}. Therefore,
it is fair to say that the idea for the self-tuning has
been suggested in the RS type models but there has not appeared
a successful model so far.

In this Letter, we present a working self-tuning model
with $1/H^2$ term in the Lagrangian in the 5D RSII 
model~\cite{rs2}. The basic mechanism that our solution works is
the following. The solution with a 4D flat space ansatz is 
regular at the whole $y$ space and introduces two integration 
constants $a$ and $c$. One constant $a$ defines the Planck scale.
The other integration constant $c$ participates in the
boundary condition at B1 and is related with the brane tension
and bulk cosmological constant.[Note that in the RSII model, there
appears one integration constant which is not participating
in the boundary condition.]

We proceed to discuss the model with the action,
\begin{eqnarray}
S&=&\int d^4x\int dy \sqrt{-g}\bigg(\frac{1}{2}R
+\frac{2\cdot 4!}{H_{MNPQ}H^{MNPQ}}\nonumber \\
&-&\Lambda_b+{\cal L}_m\delta(y)
\bigg)\label{action}
\end{eqnarray}
where we put the brane B1 at $y=0$ and the brane tension at B1 is
$\Lambda_1\equiv -\langle{\cal L}_m\rangle$ . 
We set the fundamental mass parameter
$M$ as 1 and we will recover the mass $M$ wherever it is explicitly 
needed. We assume a $Z_2$ symmetry of the solution, 
$\beta(-y)=\beta(y)$. We introduced the three index antisymmetric 
tensor field $A_{MNP}$ whose field strength is denoted as $H_{MNPQ}$.
The action contains the $1/H^2$ term which does not make sense if
$H^2$ does not develop a vacuum expectation value. We anticipate
that this term constitutes a part of the gravitational interactions,
and hence the renormalizability is not considered in this paper. If
there results a good solution for the cosmological constant problem,
it can be more seriously considered as a fundamental interaction.

The ansatz for the metric is taken as~\cite{curved}
\begin{equation}
ds^2=\beta^2(y)\eta_{\mu\nu}dx^\mu dx^\nu+dy^2 \label{metric}
\end{equation}
where $(\eta_{\mu\nu})={\rm diag.}(-1,+1,+1,+1)$. 
With the brane tension $\Lambda_1$ at B1 and the bulk cosmological 
constant $\Lambda_b$, the energy momentum tensors are 
\begin{eqnarray}
T_{MN}&=& -g_{MN}\Lambda_b-g_{\mu\nu}\delta_M^\mu\delta_N^\nu
\Lambda_1 \delta(y)\nonumber \\
&+&4\cdot 4!\left(\frac{4}{H^4}H_{MPQR}H_N\,^{PQR}
+\frac{1}{2}g_{MN}\frac{1}{H^2}\right).\label{emtensor}
\end{eqnarray}

$H_{MNPQ}$ has been considered before in connection with the
cosmological constant problem~\cite{witten} and 
compactification~\cite{fr}. The specific form for $H^2\equiv H_{MNPQ}
H^{MNPQ}$ in Eq.~(\ref{action}) 
makes sense only if $H^2$ develops a vacuum expectation
value at the order of the fundamental mass scale. Because of the
gauge invariant four index $H_{MNPQ}$, four space-time is 
singled out from the five dimensions~\cite{fr}. 
The ansatz for the four form field is 
$H_{\mu\nu\rho\sigma} 
=\sqrt{-g}{\epsilon_{\mu\nu\rho\sigma}}/{n(y)}$,
where $\mu,\cdots$ run over the Minkowski indices 0, 1, 2, and 3.
In the 5D space, the three index antisymmetric tensor
field is basically a scalar field $a$ defined by
$\p_M a=(1/4!)\sqrt{-g}\epsilon_{MNPQR}H^{NPQR}$. 

In this Letter, we show that there exists a solution for 
$\Lambda_b < 0$.  The two relevant Einstein equations are the (55) 
and ($\mu\mu$) components,
\begin{eqnarray}
6\left(\frac{\beta^\prime}{\beta}\right)^2&=&-\Lambda_b
- \frac{\beta^8}{A}\label{eqnb}\\
3\left(\frac{\beta^\prime}{\beta}\right)^2+3\left(
\frac{\beta^{\prime\prime}}{\beta}\right)&=&
-\Lambda_b-\Lambda_1\delta(y)-3\frac{\beta^8}{A}\label{eqnb1}
\end{eqnarray}
where prime denotes the derivative with respect to $y$ and $A$ is 
a positive constant in view of the ansatz of $H_{\mu\nu\rho\sigma}$.
It is easy to check that Eq.~(\ref{eqnb1}) in the bulk is obtained 
from Eq.~(\ref{eqnb}) for any $\Lambda_b,\Lambda_1,$ and $A$. 
If we took $H^2$ (instead of $1/H^2$ term) in the
Lagrangian, this statement will still hold but the resulting 
solutions do not lead to a self-tuning solution~\cite{kkl2}. 
If we had both $1/H^2$ and $H^2$ terms, there does not result a self-tuning
solution.
Near B1(the $y=0$ brane), the $\delta$ function must be
generated by the second derivative of $\beta$. The
$Z_2$ symmetry, $\beta(-y)=\beta(y)$, implies
$(d/dy)\beta(y)|_{0^+}=-(d/dy)\beta(y)|_{0^-}$. Thus,
we can write
$({d^2}/{dy^2})\beta(|y|)$ as $({d^2}/{dy^2})\beta(|y|)|_{
y\ne 0}+2\delta(y)({d}/{d|y|})\beta(|y|)$.
This $\delta$-function condition at B1 leads to a boundary condition
\begin{equation}
\frac{\beta^\prime}{\beta}\Big|_{y=0^+}\equiv -k_1,\label{bc1}
\end{equation}
where we define $k$'s in terms of the bulk cosmological constant
and the brane tension,
\begin{equation}
k\equiv\sqrt{-\frac{\Lambda_b}{6}},\ \ \ 
k_1\equiv \frac{\Lambda_1}{6}.\label{k}
\end{equation}
It is sufficient if we find a solution for the bulk equation
Eq.~(\ref{eqnb}) with the boundary condition Eq.~(\ref{bc1}). 
We define $a$ in terms of $A$,
\begin{equation}
a=\sqrt{\frac{1}{6A}}.\label{a}
\end{equation}
We note that the solution $\beta(y)$ should satisfy:\\
\indent (i) the metric is well-behaved in the whole region
of the bulk, and\\
\indent (ii) the resulting 4D effective Planck mass is finite.\\

The solution of Eq.~(\ref{eqnb}) consistent with the $Z_2$ symmetry is
\begin{equation}
\beta(|y|)=\left(\frac{k}{a}\right)^{1/4} {[\cosh(4k|y|+c)]^{-1/4}},
\label{solution}
\end{equation}
where $c$ is an integration constant to be determined by the 
boundary condition Eq.~(\ref{bc1}). This solution, consistent
with (i), is possible for a finite range of
the brane tension $\Lambda_1$. Note that $c$ can take any sign.
This solution gives a localized
gravity consistent with the above condition (ii).
The boundary condition (\ref{bc1}) determines $c$ in terms of 
$\Lambda_b$ and $\Lambda_1$,
\begin{equation}
c=\tanh^{-1}\left(\frac{k_1}{k}\right)=\tanh^{-1}
\left(\frac{\Lambda_1}{\sqrt{-6\Lambda_b}}\right).\label{c}
\end{equation}
We note that the solution exists for a finite range of
parameters, $-\sqrt{-6\Lambda_b}<\Lambda_1<\sqrt{-6\Lambda_b}$.

The effective 4D Planck mass is finite
\begin{equation}
M_{P,\rm eff}^2=2M^3\left(\frac{k}{a}\right)^{1/2}\int_0^\infty
dy\frac{1}{\sqrt{\cosh(4ky+c)}}.\label{planck}
\end{equation}
 
Note that the Planck 
mass is given in terms of the integration constant $a$, or 
the integration constant is expressed in
terms of the fundamental mass $M$ and the 4D Planck mass $M_{P,\rm eff}$,
\begin{equation}
a=\left(\frac{M^3}{M_{P,\rm eff}^2\sqrt{2k}}F\right)^2.\label{bca}\label{A}
\end{equation}
where $F$ is an elliptic integral of the first kind
and is known to be finite~\cite{kkl2}. 

To obtain the field equation for $A_{MNP}$, we note that
the variation of the Lagrangian (\ref{action}) with respect to
$A_{NPQ}$ gives
\begin{eqnarray}
\delta {\cal L} &\supset& -4\cdot 4!\ \p_M\left(\sqrt{-g}
\frac{H^{MNPQ}\delta A_{NPQ}}{H^4}\right)\nonumber\\
&+&4\cdot 4!\left[\p_M\left(\sqrt{-g}\frac{H^{MNPQ}}{H^4}\right)\right]
\delta A_{NPQ} \label{L5}.
\end{eqnarray}
To cancel the first term of the above equation, we add a surface 
term in the action
$$
S_{\rm surface}=\int d^4xdy \ 4\cdot 4!\ \p_M\left(
\sqrt{-g}\frac{H^{MNPQ}A_{NPQ}}{H^4}\right)
$$
\begin{equation}
=\int d^4xdy\ 
4\cdot 4!\left[\frac{\sqrt{-g}}{H^2}+A_{NPQ}\p_M
\left(\sqrt{-g}\frac{H^{MNPQ}}{H^4}\right)\right]\label{surface}
\end{equation}
where the variation of derivative of $A_{NPQ}$ vanishes at
the boundary~\cite{duncan}.
Then the field equation for $A_{NPQ}$ is
\begin{equation}
\p_M \frac{\sqrt{-g}H^{MNPQ}}{H^4}=0,\label{heqn}
\end{equation} 
which can be integrated to give
\begin{equation}
\frac{\sqrt{-g}H^{MNPQ}}{H^4}
={\rm function\ of\ } y\ {\rm only}.
\end{equation}
Due to our ansatz for the 4D homogeneous space, 
$H_{MNPQ}$ can have nonvanishing values only for $H_{\mu\nu\rho\sigma}$
as discussed before. Thus, $A$ in Eqs.~(\ref{eqnb}) and 
(\ref{eqnb1}) and hence $a$ in Eq.~(\ref{a}) is an integration 
constant. Field equations do not determine $a$, namely $a$ is not 
dynamically determined. But $a$ can take any value. Then for a given 
$a$, the Planck mass is given in terms of $a$ as shown in 
Eq.~(\ref{planck}). It is clear that this integration constant $a$ 
itself does not participate in the self-tuning. On the other hand, 
the integration constant $c$ participates in the self-tuning. 

Suppose we are given with $\Lambda_1$ and $\Lambda_b$. Then 
we can always find a solution for $\Lambda_b<0$ and 
$|\Lambda_1|<\sqrt{-6\Lambda_b}$. 
Namely, there exists a 4D flat space solution 
(\ref{metric}) with $c$ given by Eq.~(\ref{c}). If we add some 
constant vacuum energy at B1, then $\Lambda_1$ is shifted to say 
$\Lambda_1^\prime$.  For this new set of $\Lambda_1^\prime$ and 
$\Lambda_b$, again we can find a solution, but with a different 
integration constant $c^\prime$ given with $\Lambda_1^\prime$ 
through Eq.~(\ref{c}). In other words, the dynamics of gravity and 
the antisymmetric tensor field adjust solutions a little bit, i.e. 
self-tune the above integration constant from $c$ to $c^\prime$, to 
satisfy the field equations.

Even though we obtained a flat space solution for the 4D Minkowski
space, it is worthwhile to check explicitly that the effective
cosmological constant vanishes.  From the action 
(\ref{action}), the 4D gravity with vacuum energy 
is effectively described by
\begin{eqnarray}
S&=&\int d^4xdy\sqrt{-\eta}\beta^4
\bigg[\frac{1}{2}\beta^{-2}\tilde R_4-
4\left(\frac{\beta^{\prime\prime}}{\beta}\right)-6\left(
\frac{\beta^\prime}{\beta}\right)^2\nonumber \\
&-&\Lambda_b +\frac{2\cdot 4!}{H^2}-\Lambda_1\delta(y)\bigg]+S_{\rm surface}
\label{cci}
\end{eqnarray}
where the 4D metric is $\tilde g_{\mu\nu}
=\beta^2\eta_{\mu\nu}$, $\eta$ is the determinant of 
$\eta_{\mu\nu}$, and $\tilde R_4$ is the 4D Ricci scalar.
Then $-\Lambda_{\rm eff}$ is given by the $y$ integral of
Eq.~(\ref{cci}) except the $\tilde R_4$ term,
\begin{eqnarray}
\Lambda_{\rm eff}&=&\int_{-\infty}^{\infty} dy\ \beta^4
\bigg[4\left(\frac{\beta^{\prime\prime}}{\beta}\right)+6
\left(\frac{\beta^\prime}{\beta} \right)^2+\Lambda_b\nonumber\\
&+&\frac{\beta^8}{A} + \Lambda_1\delta(y) +\frac{2\beta^8}{A}\bigg].
\end{eqnarray}
Using Eqs.~(\ref{eqnb}) and (\ref{eqnb1}), we can rewrite
$\Lambda_{\rm eff}$ as
\begin{equation}
\Lambda_{\rm eff}=\Lambda_{\rm eff}^{(1)}+
\Lambda_{\rm eff}^{(2)},
\end{equation}
where
\begin{eqnarray}
\Lambda_{\rm eff}^{(1)}&=&-\int_{-\infty}^\infty dy\left(
\frac{2}{3}\Lambda_b+\frac{1}{3}\Lambda_1\delta(y)\right)\beta^4,
\nonumber\\ 
\Lambda_{\rm eff}^{(2)}&=&-\frac{8}{3A}\int_0^\infty dy
\beta^{12}.
\end{eqnarray}
Using the solution (\ref{solution}), and conditions (\ref{k}),
(\ref{a}) and (\ref{c}), we can show that 
\begin{eqnarray}
\Lambda_{\rm eff}^{(1)}&=&-2 \frac{kk_1}{a}\frac{1}{\cosh(c)}
+\left[2\frac{k^2}{a}\tan^{-1}\sinh(4ky+c)\right]_0^\infty
\nonumber\\
\Lambda_{\rm eff}^{(2)}&=&2\frac{k^2}{a}\frac{\sinh(c)}{\cosh^2(c)}
-\left[2\frac{k^2}{a}\tan^{-1}\sinh(4ky+c)\right]_0^\infty,
\end{eqnarray}
which leads to $\Lambda_{\rm eff}^{(1)}+\Lambda_{\rm eff}^{(2)}=0$,
in agreement with the flat 4D metric, Eq.~(\ref{metric}).

So far we presented a model for a simple Lagrangian with
$1/H^2$ term only. However, we can show that more general Lagrangians 
containing only negative powers of $H^2$ can have the self-tuning
solutions. If the Lagrangian contains 
$\sum_{n}(a_n /(\langle H^2\rangle)^n)$
with $a_1>0$ and large for the $1/\langle H^2 \rangle$ dominance, 
the last terms of the (55) and $(\mu\mu)$ Einstein 
equations, (\ref{eqnb}) and (\ref{eqnb1}), are changed to 
\begin{eqnarray}
-\sum_{n}\frac{C_n(\beta)}{A_n},\ \ \ 
-\sum_{n}\frac{(2n+1)C_n(\beta)}{A_n},
\end{eqnarray}
respectively with $A_1>0$. Then checking the two equations, we obtain
$C_n(\beta)=\beta^{8n}$ for the consistency. Then, the (55) equation 
gives
\begin{equation}
|\beta^\prime|=\sqrt{-\frac{\Lambda_b}{6}\beta^2-\sum_n
\frac{\beta^{8n+2}}{6A_n}}.\label{general}
\end{equation} 
Eq. (\ref{general}) gives $\beta^\prime\rightarrow 0$
as $\beta\rightarrow 0$ if $n\ge 0$, which guarantees
the existence of the solution. But for $n<0$,
there exists a naked singularity and there is no solution.
Suppose that the $1/\langle H^2\rangle$ is given and other corrections
are powers of $H^2$. Then, these small corrections 
in $(H^2)^m$ ($m>0$) can be brought approximately into the form
$\sum_{n\ge 0}a_n/(\langle H^2\rangle+H^2)^n$ 
where the $n=1$ term is
dominant. Thus, if the corrections contain only the $(H^2)^n$ type
terms, the existence of the self-tuning solution is intact.

Before concluding, we point out that the self-tuning solution
Eq.~(\ref{solution}) is stable in the sense that the
metric perturbation around the solution does not lead to
tachyons~\cite{kkl2}. 

In conclusion, we obtained a solution for self-tuning of the
cosmological constant in the 5D theory with the $Z_2$ symmetry.
For the self-tuning solution to exist, the bulk cosmological constant 
must be negative, $\Lambda_b<0$, and we need a specific form for
the gravitational interaction of the three index antisymmetric 
tensor field $A_{MNP}$.

\acknowledgments
This work is supported in part by the BK21 program of Ministry 
of Education, Korea Research Foundation Grant No. KRF-2000-015-DP0072, 
CTP Research Fund of Seoul National University,
and by the Center for High Energy Physics(CHEP),
Kyungpook National University.


\begin{references}
\bibitem{curved} If we took the curved space ansatze, still we obtain
solutions~\cite{kkl2}. 
But as stated in the introduction, the self-tuning solution 
means the existence of the flat space solution.
\end{references}
\end{document}